\documentstyle[prb,aps,ifthen,twocolumn]{revtex}
\input epsf

\newcommand{\eqn}[1]{Eq. (#1)}
\newcommand{\SPIN}{{\mathbf S}}
\newcommand{\orbit}{O}
\newcommand{\zee}{{\tilde z}}

\begin{document}

\title{Semiclassical mechanics of a non-integrable spin cluster}
\author{P.~A.~Houle\cite{houlenote}, 
N.~G.~Zhang, and C.~L.~Henley\cite{clhnote}}
\address{Laboratory of Atomic and Solid State Physics, Cornell University, Ithaca, NY, 14853-2501}
\maketitle

\begin{abstract} 
We study detailed classical-quantum correspondence for 
a cluster system of three spins with single-axis anisotropic exchange coupling.
With autoregressive spectral estimation,  we find oscillating terms in the
quantum density of states caused by classical periodic orbits:  in the slowly
varying part of the density of states we see signs of nontrivial topology
changes happening to the energy surface as the energy is varied.  Also,
we can explain the hierarchy of quantum energy levels near the ferromagnetic
and antiferromagnetic states with EKB quantization to explain large structures
and tunneling to explain small structures.
\end{abstract}

  %%% Classical spin models
%%% NO 75.45.+j Macroscopic quantum phen. in magnetic systems

%%% Local Variables: 
%%% mode: latex
%%% TeX-master: "document"
%%% End: 
\narrowtext

%%% Local Variables: 
%%% mode: latex
%%% TeX-master: "document"
%%% End: 

\section{Introduction}

When $S$ is large,  spin systems can be modeled by classical and
semiclassical techniques.  
Here we reserve ``semiclassical'' to mean not
only that the technique works
in the limit of large $S$ (as the term is sometimes used)
but that it implements the quantum-classical correspondence 
(relating classical trajectories to quantum-mechanical behavior).

Spin systems (in particular $S=1/2$) are often thought 
%% Although,  motivated by the case of $S=1/2,$  some think
as the antithesis of the classical limit.
Notwithstanding that, classical-quantum correspondence
has been studied at large values of $S$
in systems such as an autonomous single spin \cite{Shankar80},
kicked single spin \cite{Haake87},  and autonomous two \cite{Srivastava90c} and
three \cite{Nakamura93} spin systems.  

%%% in terms of WKB quantization and dependence

When the classical motion has a chaotic regime,  for example, 
the dependence of level statistics on the regularity of classical motion 
has been studied \cite{Srivastava90c,Nakamura93}. 
In regimes where the motion is predominantly regular,  the pattern
of quantum levels of a spin cluster can be understood with a combination of EBK (Einstein-Brillouin-Keller, also called 
Bohr-Sommerfeld) quantization
and tunnel splitting (Sec.~\ref{sec-clustering} is a such a study for the current
system.)  
The latter sort of calculation has potential applications to some
problems of current numerical or experimental interest. 
Numerical diagonalizations for extended spin systems (in ordered phases)
on lattices of modest size ($10$ to $36$ spins) may be analyzed
by treating the net spin of each sublattice as a single large spin and
thereby reducing the system to an autonomous cluster of a few spins;
the clustering of low-lying eigenvalues can probe symmetry breakings
that are obscured in a system of such size if only ground-state correlations are
examined.~\cite{Henley98}
Nonlinear self-localized modes in spin lattices
\cite{Lai97a}, which typically span several sites, 
have to date been modeled classically, but seem well suited to 
semiclassical techniques. 
Another topic of recent experiments is
the molecular magnets~\cite{Gatteschi94}
such as $\rm Mn_{12}Ac$ and $\rm Fe_8$, 
which are more precisely modeled
as clusters of several interacting spins rather than a single
large spin;
%%% as the number of eigenstates increases
semiclassical analysis may provide an alternative to 
exact diagonalization techniques \cite{Katsnelson99}
for theoretical studies of such models.

In this paper, we will study three aspects of the classical-correspondence
of an autonomous cluster of three spins coupled by easy-plane exchange
anisotropy, with the Hamiltonian
\begin{equation}
  {H} = \left[ \sum_{i=1}^3 \SPIN_i \cdot \SPIN_{i+1} - \sigma \SPIN_i^z \SPIN_{i+1}^z \right],
\label{eq:ham3a}
\end{equation}
This model was introduced in Ref.~\onlinecite{Nakamura85},  a study of level
repulsion in regions of $(E,\sigma)$ space where the classical dynamics
is predominantly chaotic \cite{Nakamura93}.  
\eqn{\ref{eq:ham3a}} has only two nontrivial degrees of freedom,  since it
conserves total angular momentum around the z-axis.  
As did Ref.~\onlinecite{Nakamura85} we consider only the
case of $\sum_i \SPIN_i^z=0.$  While studying classical mechanics we set $|\SPIN|=1;$
to compare quantum energy levels at different $S$, we we divide energies by by $S(S+1)$ to
normalize them.  The classical maximum energy, $E=3$, occurs at the
ferromagnetic (FM) state -- all three spins are
coaligned in the equatorial (easy) plane.  
The classical ground state energy
%%% of opposite chirality 
is $E=-1.5$,  
in the antiferromagnetic (AFM) state, in which
the spins lie $120^\circ$ apart in the easy plane; there are
two such states,  differing by a reflection of the spins in 
a plane containing the $z$ axis. 
Both the FM and AFM
states,  as well as all other states of the system,  are continuously degenerate with
respect to rotations around the $z$-axis.
The classical dynamics follows from the fact that $\cos\theta_i$ and $\phi_i$, 
are conjugate, where $\theta_i$ and $\phi_i$ are
the polar angles of the unit vector $\SPIN_i$;
then Hamilton's equations of motion say 
   \begin{eqnarray}
        d\cos \theta_i/dt &=& \hbar^{-1} \partial {H} / \partial \phi_i;
        \nonumber  \\
        d\phi_i/dt &=& - \hbar^{-1} \partial {H} / \partial \phi_i.
   \end{eqnarray}

In the rest of this paper, 
we will first introduce the classical dynamics 
by surveying the fundamental periodic orbits of the three-spin cluster,  
determined by
numerical integration of the equations of motion (Sec~\ref{sec-classical}).  
The heart of the paper is Sec.~\ref{sec-orbitspectrum}: 
starting from the quantum density of states (DOS) obtained from
numerical diagonalization, 
we apply nonlinear spectral analysis 
to detect the oscillations
in the quantum DOS caused by classical periodic orbits;
to our knowledge, 
this is the first time the DOS has been related to specific
orbits in a multi-spin system. 
Also,  in Sec.~\ref{sec-flatspot} we smooth the
DOS and compare it to a lowest-order Thomas-Fermi approximation
counted by Monte Carlo integration of the classical energy surface;
a flat interval is visible in the quantum DOS
%% the smoothed density of states 
between two critical energies where
the topology of the classical energy surface changes. 
Finally, in Sec.~\ref{sec-clustering}, we use a combination of
EBK quantization and tunneling analysis to explain the clustering
patterns of the quantum levels in our system. 

\section {Classical periodic orbits}
%%% {\it Classical Periodic Orbits --}
\label{sec-classical}

\begin{figure}
  \centerline{\epsfxsize=3.375in\epsfbox{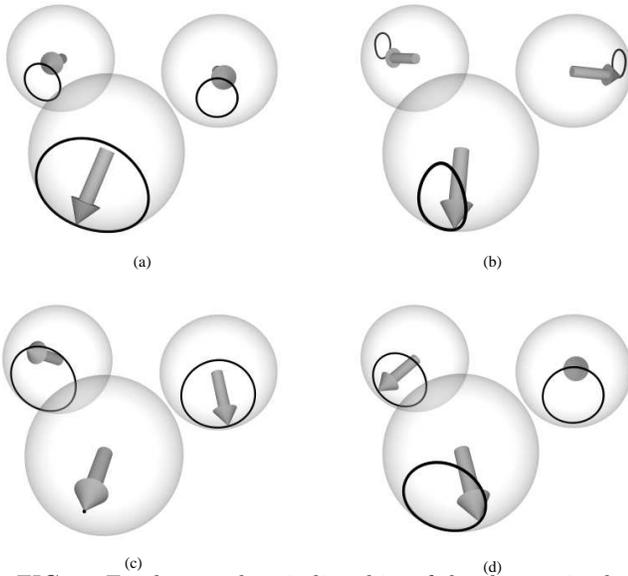}}
  \caption{Fundamental periodic orbits of the three-spin cluster.  (a) Counterbalanced,  (b) unbalanced orbits,
(c) stationary spin,  and (d) three-phase.}
  \label{fig:tracks}
\end{figure}

Our subsequent semiclassical analysis will depend on identification of
all the fundamental orbits and their qualitative changes as parameters
are varied. 
Examining Poincar\'e  sections and searching along symmetry lines of the system,
we found four families of fundamental periodic orbits
for the three-spin cluster.  Figure \ref{fig:tracks} is an illustration of their motion,
and Figure \ref{fig:classicalEtau} gives classical energy-time curves.  Orbits of types (a)-(c) are 
always at least threefold degenerate,  since one spin is different from the other two;  orbits of
types (a)-(c) are also time-reversal invariant.  Orbit (a),  the {\it counterbalanced}
orbit,  exists when $E>-1$ (including the FM limit)
and,  in the range $0<\sigma<1$ which we've studied,  
is always stable.  Orbit (b), the {\it unbalanced} orbit,  is unstable and exists when $E<E_p,$  where
\begin{equation}
E_p={3 \over 4} \sigma - {3 \over 2}. \label{eq:polar}
\end{equation}
Orbits of type (c),  or {\it stationary spin}
exist at all energies.  Type (c) orbits are are unstable in the range, $-1>E>3.$ 
Below $E=-1$ the stationary spin orbit bifurcates into two branches without
breaking the symmetry of
the ferromagnetic ground state.  At
  \begin{equation}
    E_c(\sigma)={3 - 3 \sigma + \sigma^2 \over \sigma - 2},
  \label{eq:afmcat}
  \end{equation}
one branch vanishes and the other branch bifurcates into two orbits that are distorted
spin waves of the two AFM ground states.  
(Below, in Sections~\ref{sec-flatspot} and \ref{sec-clustering},
we will discuss topology changes
of the entire energy surface.)
Although they are not related by symmetry, all  orbits of type
(c) at a particular energy have the same period.  

Orbits of type (d),  or {\it three-phase} orbits are
named in analogy to three-phase AC electricity,  as spin vectors move along distorted circles,
$120^\circ$ out of phase.  
The type (d) orbits break time-reversal symmetry and are
hence at least twofold degenerate.  A symmetry-breaking pitchfork bifurcation of the
(d) family occurs (for $\sigma=0.5$ around $E=-0.75$) at which a single stable orbit,
approaching from high energy,  bifurcates into an unstable and two stable 
{\it precessing three-phase} orbits without period doubling.  \cite{DeAguiar87}
(Strictly speaking,
the precessing three-phase orbits are not periodic orbits of the three-spin system,
since after one ``period'' the spin configuration is not the same as before,  but rather,
all three spins are rotated by the same angle around the $z$-axis).  
The unstable three-phase orbit disappears quickly as we lower energy,  
but the precessing three-phase
orbits persist until $E=-1.5,$  and become intermittently stable and unstable in a heavily
chaotic regime near $E_A,$  but regain stability before $E \rightarrow -1.5$:
thus in the AFM limit, 
orbits (c) are stable while orbits (b) are unstable.~\cite{Henon64}
More information on
the classical mechanics of this system appears in 
Refs.~\onlinecite{Houle98} and \onlinecite{Houle98c}.
%%% NOTE ADDED

\section{Orbit spectrum analysis}
\label{sec-orbitspectrum}

Gutzwiller's trace formula,  the central result of periodic orbit theory,
\cite{DeAlmedia88}
\begin{equation}
\rho(E)={\mathrm{Re}} \sum_p 
 A_p(E) \exp[i S_p(E) / \hbar]
+ \rho_{tf}(E)
, \label{eq:gtf}
\end{equation}
decomposes the quantum DOS $\rho(E)$ into a sum
of oscillating terms contributed by
classical orbits indexed by $p,$
where $S_p(E)$ is the classical action,  and
$A_p(E)$ is a slowly varying function of the period,
stability and geometric~\cite{FN-maslov}
%%% \footnote{$A_p(E)$ contains a phase factor,
%%% $e^{i\mu/4}$,  where $\mu$ is the Maslov index
%%% and depends on the topology of the linearized
%%% dynamics near the orbit,  see Ref. \onlinecite{Robbins91}.
%%% As we do not consider amplitude or phase,  the
%%% Maslov index is irrelevant for this paper.}
properties of the orbit $p$), plus 
the zeroth-order {\it Thomas-Fermi} term,
\begin{equation}
\rho_{tf}(E)=\int { d^{2N}\zee \over (2 \pi \hbar)^N } \delta \left(E-H(\zee) \right),
\label{eq:thomasFermi}
\end{equation}
This integral over phase space $\zee$ is simply
proportional to the area of the energy surface.
We do not know of any mathematical derivation of (\ref{eq:gtf})
in the case of a spin system.

At a fixed H,  the {\it orbit spectrum} is,  as function of $\tau,$
the power spectrum of $\rho(E)$
inside the energy window,  $H-\Delta H/2<E<H+\Delta H/2.$  
(Figure \ref{fig:c3os},  explained below, is an example of an orbit spectrum.)
Since the classical period $\tau_p(E)=\partial S_p(E) / \partial E,$
\eqn{\ref{eq:gtf}} implies that $\orbit(H,\tau)$ is large if there
exists a periodic orbit with energy $H$ and period $\tau.$  The orbit
spectrum can be estimated by Fourier transform,~\cite{FN-FTnote}
%%% \footnote{\eqn{\ref{eq:ftospec}} is for illustration.  The square window
%%% aggravates artifacts of the Fourier transform which can be reduced by
%%% using a different window function.\onlinecite{Marple87}}
%
\begin{equation}
\orbit(H,\tau)= 
\left|
\int_{H-\Delta H/2}^{H+\Delta H/2}
\rho(E) e^{-i \hbar^{-1} E \tau} dE \right|^2. \label{eq:ftospec}
\label{eq-fourier}
\end{equation}
Variants of \eqn{\ref{eq:ftospec}} have been used to extract
information about classical periodic orbits from quantum spectra.
\cite{Baranger95,Ezra96}
Unfortunately,  the resolution of the
Fourier transform is limited by the uncertainty principle,
$\delta E \delta t = \hbar/2.$

Nonlinear spectral estimation techniques,  however, can surpass the 
resolution of the Fourier transform.  \cite{Marple87}
One such technique,  harmonic inversion,  has been successfully
applied to scaling systems \cite{Main97} --
i.e., systems like billiards or Kepler systems
in which the (classical and quantum) dynamics at one energy are 
identical to those at any other energy, after a rescaling of 
time and coordinate scales. 
In a scaling system, 
windowing is unnecessary because there are no bifurcations and
the scaled periods of orbits are constant.
In this section,  we will apply nonlinear spectral
estimation to our system (\ref{eq:ham3a}), 
which is {\it nonscaling}.~\cite{FN-scaling}
%%% \footnote{In fact, the {\it classical dynamics} of our system do scale 
%%% if one rescales the spin length $S$ simultaneously. 
%%% However,  in contrast to the mentioned scaling systems systems,  the
%%% $S$ is a spin system is not just a numerical parameter, but is a discrete quantum number. 
%%% In effect, constructing an orbit spectrum by varying $S$ is a mixture of
%%% scaled-energy spectroscopy and inverse-$\hbar$ spectroscopy 
%%% \cite{Main97c}.   
%%% Such an approach would give poor energy resolution in our system, since we can perform diagonalizations
%%% only for discrete values of $S$ in a limited range.}

\subsection {Diagonalization}
\label{subsec-diagonalize}
%%% {\it Diagonalization -- }
To get the quantum level spectrum,  we wrote software to
diagonalize arbitrary spin Hamiltonians  polynomial in ($\SPIN_i^x,\SPIN_i^y,\SPIN_i^z$), 
where $i$ is an index running over arbitrary $N$ spins of arbitrary (and often large) spin
$S$. The program,  written in {\tt Java}, takes advantage of discrete translational
and parity symmetries by constructing a basis set in which the Hamiltonian is block
diagonal,  letting us diagonalize the blocks independently with an optimized
version of LAPACK.  Picturing the spins in a ring,  the Hamiltonian 
\eqn{\ref{eq:ham3a}} is invariant to cyclic permutations of the spins,
so the eigenstates are states of definite wavenumber~\cite{Nakamura93}
$k=0,\pm { 2\pi \over 3}$ 
(matrix blocks for
$k= \pm { 2 \pi \over 3}$ are identical by symmetry). 
In the largest system we diagonalized (three-spin cluster with $S=65$) ,
the largest blocks contained $N=4620$ states.   

%%% {\it Orbit spectrum -- }
\subsection{Autoregressive approach to construct spectrum}

The input to an orbit spectrum calculation is the list of discrete eigenenergies with total $S_z=0$;
no other information on the eigenstates (e.g. the wavenumber quantum number)
is necessary. 
This level spectrum is smoothed by convolving with a
Gaussian (width $10^{-3}$ for 
Figure \ref{fig:c3os}) and discretely sampling over energy
(with sample spacing $\delta=4.5 \times 10^{-4}$.)

We estimate the power spectrum by the autoregressive (AR)
method.  AR models a discretely sampled input signal,  $y_i$ 
(in our case the density of states) with a process that attempts
to predict $y_i$ from its previous values,
\begin{equation}
y_i=\sum_{j=1}^N a_i y_{i-j} + x_i. \label{eq:filter}
\end{equation}
Here $N$ is a free parameter which
determines how many spectral peaks that model can fit;
Refs.~\onlinecite{Marple87} and
\onlinecite{Wu97} discuss guidelines for choosing $N$.
Fast algorithms exist to implement least-squares, i.e. 
to choose $N$ coefficients  $a_i$ to minimize 
(within constraints) $\sum x_i^2$; 
of these we used the Burg algorithm \cite{Marple87}. 

To estimate the power spectrum,  we discard the original $x_i$ and
model $x_i$ with uncorrelated white noise.  Thinking of  \eqn{\ref{eq:filter}}
as a filter acting on $x_i,$  the power spectrum of $y_i$ is computed
from the transfer function of \eqn{\ref{eq:filter}} and is
\begin{equation}
P(\nu)={<x_i^2> \over 1 - \sum_{j=1}^N a_j e^{i \nu \delta}}.
\end{equation}
Unlike
the discrete Fourier transform,  $P(\nu)$ can be evaluated at
any value of $\nu.$
In our application, of course, $\delta$ has units of energy, so
$\nu$ (more exactly $\nu/\hbar$) actually has units of time
and is to be identified with $\tau$ in (\ref{eq-fourier}). 

\subsection{Orbit spectrum results and discussion}

\begin{figure}
  \centerline{\epsfxsize=3.375in\epsfbox{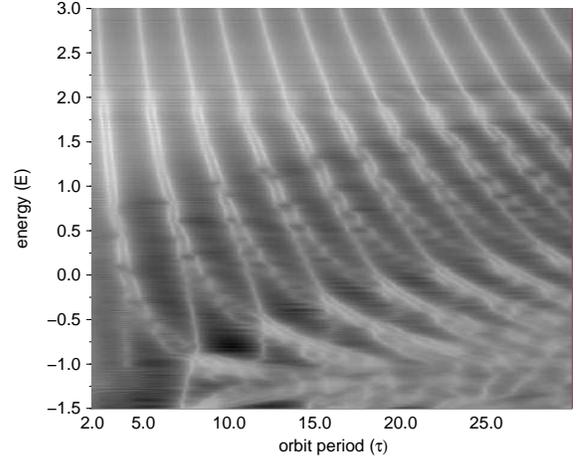}}
  \caption{Orbit spectrum for $S=65, \sigma=0.5.$  The horizontal axis is classical period, and the vertical
    axis is energy.  Peaks of the orbit spectrum are white and valleys are black.}
  \label{fig:c3os}
\end{figure}

Figure \ref{fig:c3os} shows the orbit spectrum of our system with
$S=65$ and $\sigma=0.5.$; it is displayed
as a $500 \times 390$
array of pixels,  colored light where
$\orbit(H,\tau)$ is large.  Each horizontal row is the power
spectrum in an energy window centered at $H;$  we stack rows 
of varying $H$ vertically.  With a window width 250 energy samples
long ($\delta H = 0.1125,$)  we fit $N=150$ coefficients in
\eqn{\ref{eq:filter}}.  To improve visual resolution,  we
let windows overlap and spaced the centers of successive windows 
25 samples apart.

\begin{figure}
  \centerline{\epsfxsize=2.982in\epsfbox{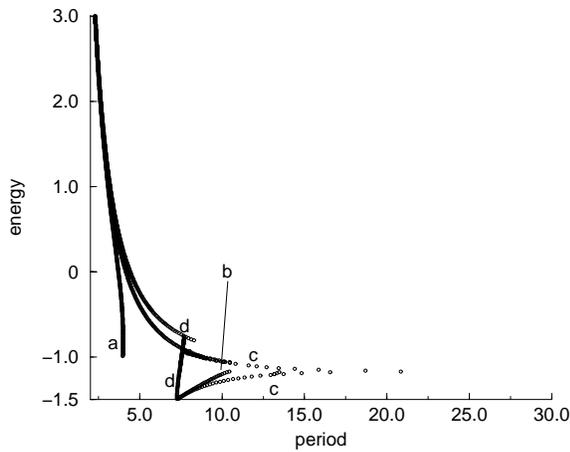}}
  \caption{
     Energy-period curve of the three spin system with $\sigma=0.5$.
The
     curves (a), (b), (c) and (d) are four families of periodic orbits
.
Below the obvious bifurcation around $E=-0.75$, orbit (d) is not
%%% (strictly speaking)
literally a periodic orbit of the three spins, but only
in a reduced two degree of freedom system (wherein one
identifies states related by a rotation of all three spins
about the $z$ axis.}
  \label{fig:classicalEtau}
\end{figure}

Comparing Figure \ref{fig:c3os} and Figure \ref{fig:classicalEtau}
we see that our orbit spectrum detects the fundamental periodic
orbits as well as multiple transversals of the orbits.  Interestingly,  we produced
Figure \ref{fig:c3os} before we had identified most of the fundamental
orbits;  Figure \ref{fig:c3os} correctly predicted three out of
four families of orbits.

%%% NOTE CHECK THIS
We believe that, given the same data, the AR method normally produces a 
far sharper spectrum. This is not surprising, since the Fourier analysis
allows the possibility of orbit-spectrum density at all $\tau$ values, 
whereas AR takes advantage of our {\em a priori} knowledge that there are
only a few fundamental periodic orbits and hence only a few peaks. 
We have compared the Fourier and AR versions of the spectrum 
in a few cases, but have not systematically tested them against each other. 

Unfortunately,  the artifacts and limitations of the AR method are
less understood than those of the Fourier transform.  At high energies,
the classical periods are nearly degenerate,  so we expect
closely spaced spectral peaks in the orbit spectrum.
In this situation,  the Burg algorithm vacillates between fitting one or two peaks
causing the braiding between the (a) and (c) orbits (labeled in
Figure \ref{fig:classicalEtau}) 
in Figure \ref{fig:c3os}.  Also,  in the range $-1<E<-1.3,$ where
classical chaos is widespread,  bifurcations increase the number of 
contributing orbits  so that we cannot interpret the orbit spectrum for $\tau>10.$

\section {Averaged density of states}
%%% {\it Averaged density of states --}
\label{sec-flatspot}

\begin{figure}
  \centerline{\epsfxsize=3.375in\epsfbox{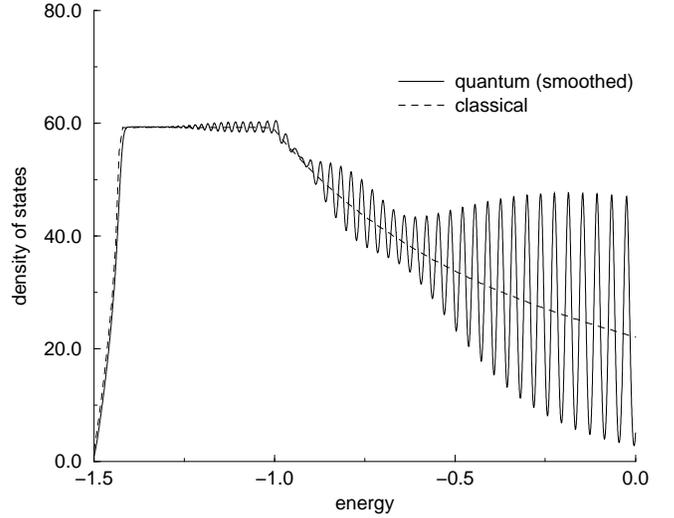}}
  \caption{
    The density of states of a $S=40$ three spin cluster
    with $\sigma=0.1,$  smoothed with a Gaussian $\exp [-
      ({E /\delta})^2 ]$ with $\delta=0.01.$
    The Thomas-Fermi density of states is very flat in the
    range $E_p<E<-1,$  with $E_p=-1.43.$ } 
  \label{fig:flatspot}
\end{figure}

The lowest-order Thomas Fermi approximation,  \eqn{\ref{eq:thomasFermi}} 
predicts that the area of the classical energy surface is proportional to
the DOS.  We verify this in Figure~\ref{fig:flatspot},  a comparison of
the heavily smoothed quantum DOS to the area of the energy surface
computed by Monte Carlo integration.  

An energy interval is visible in which the quantum DOS appears to
be constant; we then verified that the
classical DOS (which is more precise) is constant to our numerical precision; 
a similar interval was observed for all values of $\sigma$. 
We identified this interval as $(E_p, -1)$, 
where the endpoints are associated with 
changes in the topology of the energy surface as the energy varies. 

At energies below
$E_c$ (see \eqn{\ref{eq:afmcat}}),  the energy surface consists of two
disconnected pieces,  one surrounding each AFM ground state.  The two
parts coalesce as the energy surface becomes multiply connected at $E_c.$ 
For $E< E_p,$  (see \eqn{\ref{eq:polar}}) the anisotropic interaction
confines the spins to a limited band of latitude away from the poles.  At 
$E_p$ it becomes possible for spins to pass over the poles.  At $E=-1,$
the holes that appeared in the energy surface at $E_c$ close up.  A
discontinuity in the slope of the area of the energy surface occurs
at energy $E_c$ (not visible in Figure \ref{fig:flatspot});  
in the range $E_p<E<-1$ the area of the energy surface (and hence the slowly
varying part of the DOS) 
seems to be constant as a function of energy. 

In the special isotropic ($\sigma=0$) case, the flat interval is
$(-1.5,-1)$  and it can be analytically derived that the DOS
is constant there. 
This is simplest for the  
smoothed quantum DOS, since for
$n=1,2,\ldots$ there are clusters of $n$ energy levels with level
spacing proportional to $n$. (A derivation also exists for
the classical case, but is less direct.) 
We have no analytic results for general $\sigma$.

This flat interval is specific to our three-spin cluster,  but we expect that
the compactness of spin phase space will,  generally,  cause changes in
the  energy surface topology of spin systems that do not occur
in traditionally studied particle systems.

\section{Level clustering}
\label{sec-clustering}

\begin{figure}
  \centerline{\epsfxsize=3.375in\epsfbox{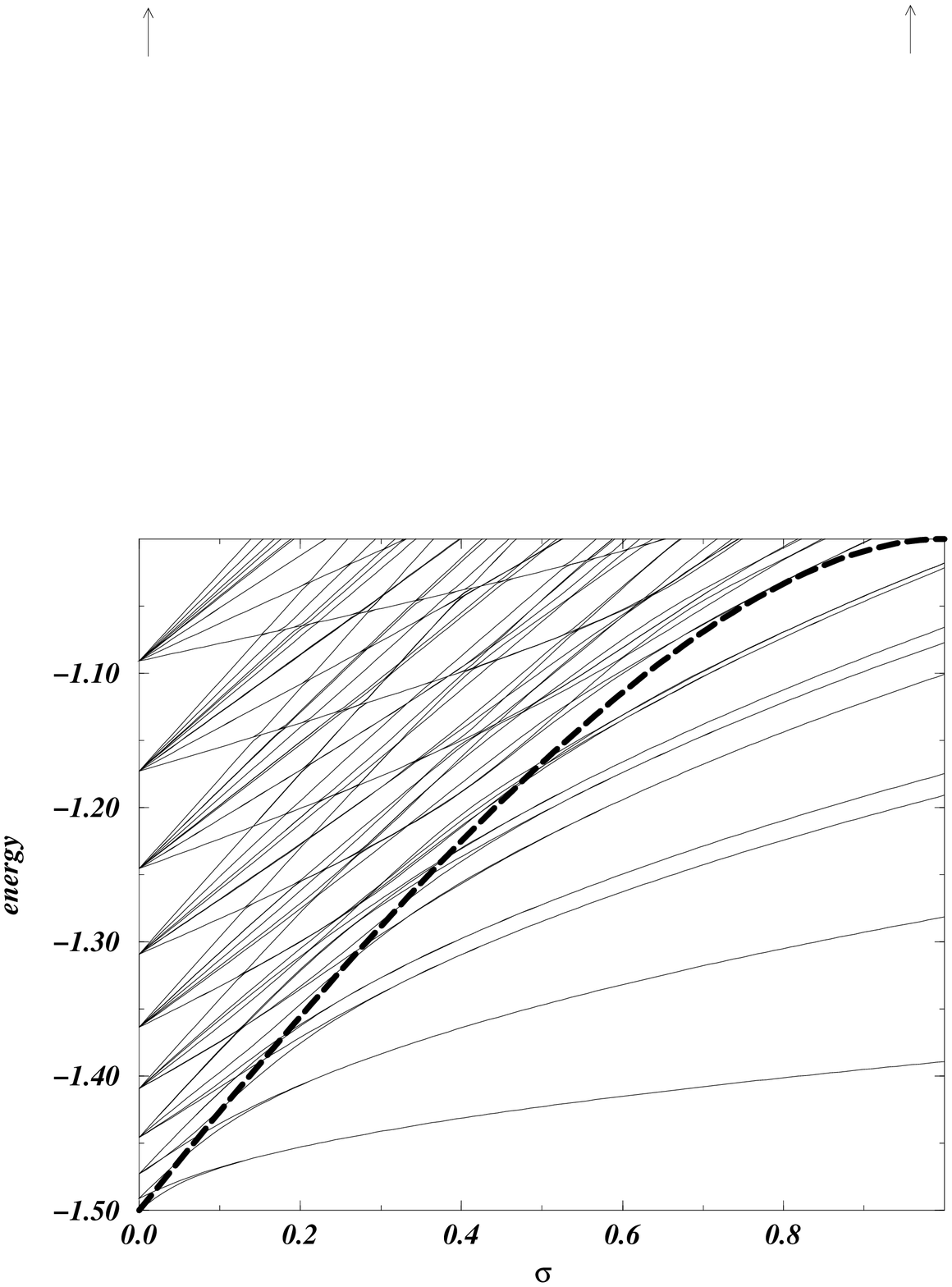}}
  \caption{
    Energy levels as a function of $\sigma$ for an
    $S=10$ three spin cluster.
    The dashed line is the 
    coalescence energy $E_c(\sigma)$ defined by
    equation \eqn{\ref{eq:afmcat}}.}
  \label{fig:c3low}
\end{figure}

The quantum levels with total $\SPIN_z=0$ show rich 
patterns of clustering,  some of which are visible on
Figure~\ref{fig:c3low}.
The levels that form clusters correspond to three different 
regimes of the classical dynamics in which the motion
becomes nearly regular: 
(1) the FM limit (not visible in Figure~\ref{fig:c3low};
(2) the AFM limit (bottom edge of Figure~\ref{fig:c3low})
and (3) the isotropic limit $\sigma=0$ (left edge of
Figure~\ref{fig:c3low}). 
Indeed, the levels form a hierarchy in as the clusters 
break up into subclusters.
In this section, we first
approximately map the phase space from four coordinates to two
coordinates -- with the topology of a {\it sphere}. 
(Two of the original six coordinates are trivial, or decoupled, 
due to symmetry, as noted in Sec.~\ref{sec-classical}.
Then, using Einstein-Brillouin-Kramers (EBK) quantization 
some consideration of quantum tunneling, 
many features of the level hierarchy will be understood.

%% CLH clipped Paul's terse version in favor of the full story
%% Near the FM and AFM classical states,  the classical
%% motion can be understood in terms of two coupled oscillators with
%% actions $I_1$ and $I_2.$  
%% The two three-phase orbits are the case where $I_1=0$ and $I_2=0.$  
%% Since the frequencies of the oscillators
%% are nearly degenerate with a 1:1 ratio,  the levels are grouped into
%% {\it clusters} which are polyads \cite{Kellman} of constant action $I=I_1+I_2.$

%% The levels within each cluster can in turn be divided into {\it sub-clusters}
%% which can be found by applying Bohr-Sommerfeld quantization to the action $I_1.$  

\subsection {Generic behavior: the polyad phase sphere}
~\label{subsec:polyads}

%%% The FM, AFM, and isotropic limits
%%% the upper, lower, and left edges of Figure~\ref{fig:c3low}  --
%%% are three distinct cases in which the motion is nearly
In all three limiting regimes, the classical dynamics becomes trivial.
For small deviations from the limit, 
the equations of motion can be linearized and 
one finds that the trajectory decomposes 
into a linear combination of 
two harmonic oscillators with degenerate frequency $\omega$, i.e., 
in a 1:1 resonance; the oscillators are coupled only by 
higher-order (=nonlinear) terms. 

There is a general prescription for understanding the
classical dynamics in this situation~\cite{Kellman}.
Near the limit, the low-excited levels have approximate quantum numbers
$n_{1,2}$ such that the excitation energy $\Delta E_i$ 
in oscillator $i$ is $\hbar \omega (n_{i}+1/2)$. 
(In the FM limit, regime (1), 
this difference is actually measured
{\it downwards} from the energy maximum.)
Clearly, the levels with a given total quantum 
number $P \equiv  (n_1+n_2+1)$
must have nearly degenerate energies, and thus form a
cluster of levels, which are split only by the effects (to be
considered shortly) of the anharmonic perturbation. 
A level cluster 
arising in this fashion is called a {\it polyad}~\cite{Kellman}.

To reduce the classical dynamics, make a canonical transformation to
the variables $\Phi$ and ${\bf P} \equiv (P_x, P_y, P_z)$, where
$\Phi$ is the 
mean of the oscillators' phases and $\Psi_x$ 
is their phase difference, and
  \begin{eqnarray}
  P_x & \equiv & {1\over 2} (n_1-n_2), \nonumber \\
  (P_y, P_z) & \equiv &
        2 \sqrt{(n_1+1/2)(n_2+1/2)}(\cos\Psi_x, \sin \Psi_x) , 
  \label{eq:polyad}
  \end{eqnarray}
Here $\Phi$ is the fast coordinate, with trivial dynamics $d\Phi/dt = \omega$  in
the harmonic limit. The slow coordinates $\bf P$ 
follow a trajectory confined to the ``polyad phase sphere''
$|{\bf P}| = P$,  since $\Delta E = \hbar \omega P $ is conserved
by the harmonic-order dynamics. 
The reduced dynamics on this sphere is properly
a map ${\bf P}_i \to {\bf P}_{i+1}$, 
defined by (say) the Poincar\'e section at 
$\Psi_x=0 ({\rm mod}\; 2\pi )$. 
But $d {\bf P}/dt$ contains only higher powers of the components of 
$\bf P$,  so near the harmonic (small $P$) limit, 
$|{\bf P}_{i+1}-{\bf P}_i|$ vanishes 
and the reduced dynamics becomes a flow. ~\cite{FN-poincare}
%%%%%%%%%%%
% Hence {\it any} Poincar\'e section rule gives essentially the same picture, 
% in particular those near the AFM or FM limits presented in Ref.~\onlinecite{Houle98c}.}
%%%%%%%%%%%
At the limit in which it is a flow, an effective Hamiltonian $I$ 
can be defined
so that the dynamics becomes integrable.~\cite{FN-I} 
Applying EBK quantization to the reduced dynamics on 
the polyad phase sphere gives the splitting of levels within a polyad
cluster.
(Near the harmonic limit, the energy scale of $I$ is small compared
to the splitting between polyads.)

\begin{figure}
  \centerline{\epsfxsize=2.4in\epsfbox{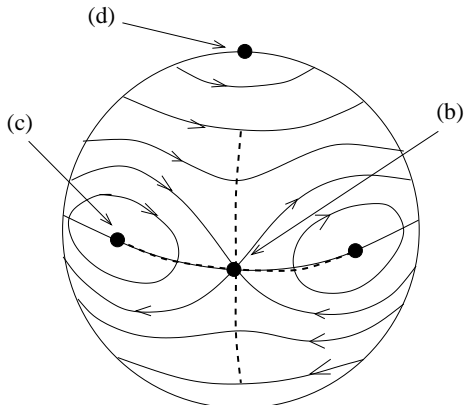}}
  \caption{
Topology of orbits for a generic phase sphere with 
3-fold symmetry. 
The phase space is decomposed into one ``fast'' action variable, 
which is transverse to the 3-space of the sphere, and 
three ``slow'' variables; of these, the energy is a function
of radius and is conserved.  
Contours are shown of an effective Hamiltonian $I$, with 
the direction of ``slow'' motion indicated by arrows. 
Tunneling takes place between symmetry-related orbits across
the saddle points of $I$, as indicated by the dashed lines.
The stationary points of this two-dimensional flow
labeled (b), (c), and (d) correspond 
near the antiferromagnetic ground state
to the periodic orbits of the full system 
labeled the same way in Figure~\ref{fig:tracks};
the labeling would differ somewhat for the 
near-ferromagnetic or 
near-isotropic regimes (see Sec.~\ref{subsec:clusterresults}).}
  \label{fig:flows}
\end{figure}

In all three of our regimes, we believe this flow has the 
topology shown schematically in Figure~\ref{fig:flows}.~\cite{FN-fourspin}
Besides reflection symmetry about the ``equator'', 
it also has a threefold rotation symmetry around the
$P_z$ axis,  which corresponds to 
the cyclic permutation of the three spins.~\cite{FN-Paxes} 
(Figure~\ref{fig:flows} is natural for the three-spin system
because it is the simplest generic topology of the phase
sphere with that threefold symmetry.)
The reduced dynamics has two symmetry-related fixed points 
at the ``poles'' $P_z=\pm P$,   which always correspond to motions 
of the three-phase sort like (d) on Figure~\ref{fig:tracks}. 
There are also three stable and three unstable fixed points 
around the ``equator''. 

The KAM tori of the full dynamics correspond to orbits of the reduced dynamics. 
These orbits follow contours of the effective Hamiltonian $I$
of the reduced dynamics
(as in Figure~\ref{fig:flows}). 
In view of the symmetries mentioned, 
   \begin{equation}
    I = \alpha P_z^2 + \beta (P_x^3 - 3 P_x P_y^2) + {\rm const}
   \label{eq:I}
   \end{equation}
 to leading order, 
where $\alpha$,$\beta$, and the constant may depend on 
$\sigma$, $S$, and $P$.

The KAM tori surrounding the three-phase orbits
represented by the ``poles''
are twofold degenerate, while  the tori in the stable resonant islands 
represented on the ``equator'' are threefold
degenerate.  
Hence, the EBK construction 
produces degenerate subclusters containing two or three levels depending
on the energy range within the polyad cluster. 

The fraction of levels in one or the other kind of subcluster
is proportional to the spherical areas on the corresponding side of the
separatrix,  which passes through the unstable points
in Figure~\ref{fig:flows}. 
These areas in turn depend on the ratio of the first to the second 
term in Eq.~(\ref{eq:I}), i.e. $\alpha P^2/ \beta P^3$. 
Evidently, as one moves away from the harmonic limit to higher
values of $P$, one universally expects to have a larger
and larger fraction of threefold subclusters. 

Given the numerical values of energy levels in a polyad, 
we can estimate the terms
of Eq.~(\ref{eq:I})
in the following fashion:
(i) the energy difference between the highest and lowest
3-fold  subcluster is the difference between the stable and
unstable orbits on the equator, which is $ 2 \beta P^3$
according to (\ref{eq:I});
(ii) the mean of the highest and lowest 3-fold subcluster
would be the energy all around the equator if $\beta$ 
were to vanish; the difference between this energy and that of
the farthest 2-fold subcluster in the polyad is $\alpha P^2$ 
according to (\ref{eq:I}).

Furthermore, tunneling between nearby tori creates {\it fine structure}
splitting inside the sub-clusters.  
The slow part of the dynamics on the polyad phase sphere, 
is identical to that of a single semiclassical spin
with (\ref{eq:I}) as its effective Hamiltonian,
so the effective Lagrangian is essentially the same, too. 
Then different tunneling paths connecting the same two
quantized orbits must differ in phase by a topological term, 
with a familiar form
proportional to the (real part of the) spherical area 
between the two paths.~\cite{Topological92}

\subsection {Results}
\label{subsec:clusterresults}

Here we summarize some observations made by
examination of polyads in the three regimes, for 
a few combinations of $S$ and $\sigma$. 

\subsubsection{Ferromagnetic limit}

This regime is the best-behaved in that 
regular behavior persists for a wide range of energies. 
The ferromagnetic state, an energy maximum, is a fixed
point of the dynamics; around it are ``spin-wave'' excitations
(viewing our system as the 3-site case of a one dimensional ferromagnet).
These are the two oscillators from which the polyad is constructed.
Thus, the ``pole'' points in Figure~\ref{fig:flows} correspond to 
``spin waves'' 
propagating clockwise or counterclockwise around the ring of three spins, 
an example of the ``three-phase'' type of orbit. 
The stable and unstable points
on the ``equator'' are identified respectively with the orbits
(a) and (c) of Figure~\ref{fig:tracks}.
Classically, in this regime, the three-phase orbit is the 
fundamental orbit with lowest frequency $\omega_{3-phase}$;
thus the corresponding
levels in successive polyads have a somewhat smaller spacing
$\hbar \omega_{3-phase}$ than other levels, and they end up at
the top of each polyad.
(Remember excitation energy is measured {\it downwards}
from the FM limit.) 
Indeed, we observe that the high-energy end of each polyad 
consists of twofold subclusters and the low-energy end 
consists of threefold subclusters.

We see a pattern of fine structure (presumably tunnel splittings)
which is just like
the pattern in the four-spin problem.\cite{Henley98,FN-fourspinclusters}
Namely, throughout each polyad
the degeneracies of successive levels follow the pattern 
(2,1,1,2) and repeat.  
%%%%%%%
(Here  -- as also for regime 3 --
every ``2''  level has $k=\pm 1$ and every ``1'' 
level has $k=0$, where wavenumber $k$ was
defined in Sec.~\ref{subsec-diagonalize}.)
%%%%%%%
Numerical data show that (independent of $S$)
the pattern (starting from the lowest energy) begins (2112...) 
for even $P$, but for odd $P$ it begins (1221...).
%%%  i.e. the opposite phase).

In the energy range of twofold subclusters, 
the levels are grouped as (2)(11)(2), i.e. 
one tunnel-split
subcluster between two unsplit subclusters(and repeat); in the 
threefold subcluster
regime, the grouping is
(21)(12), so that each subcluster gets tunnel-split into a pair and a 
single level,
but the sense of the splitting alternates from one subcluster 
to the next.

An analysis of $\sigma=0.4$, $S=30$ showed that the fraction
of threefold subclusters indeed grows from around 0.3 for small
$P$ to nearly $0.5$ at $P\approx 40$. 
Furthermore, when $\alpha P^2$ and $\beta P^3$ were estimated
by the method described near the end of Subsec.~\ref{subsec:polyads},
they indeed scaled as $P^2$ and $P^3$ respectively. 

\subsubsection {Antiferromagnetic limit}

This regime occurs at $E<E_c(\sigma)$, where $E_c(\sigma)$ is given 
by (\ref{eq:afmcat}). 
That means the classical energy surface is divided into two
disconnected pieces,  related by a mirror reflection 
of all three spins in any plane normal
to the easy plane. 
Analogous to regime one, two degenerate antiferromagnetic ``spin waves'' exist
around {\it either} energy minimum, and the polyad states are
built from the levels of these two oscillators. 
%%% chirality''$\pm 1$)   
Thus the clustering hierarchy outlined in
Sec.~\ref{subsec:polyads} -- 
polyads clusters, EBK-quantization
of $I$, and tunneling over barriers of $I$ on the polyad phase sphere --
is repeated within each disconnected piece, leading to a prediction
that {\it all} levels should be twofold degenerate. 

Consequently, on the level diagram (Figure~\ref{fig:c3low}), 
there should be half the apparent 
level density below the line $E=E_c(\sigma)$ as above it. 
Indeed, a striking qualitative change in the
apparent level crossing behavior is visible at that line
(shown dashed in the figure). 
%%% ~\ref{fig:c3low}.

Actually, {\it tunneling} is possible between the disconnected pieces
of the energy surface and may split these degenerate pairs. 
In fact this {\it hyperfine} splitting 
happens to 1/3 of the pairs, again following the
(2112) pattern within a given polyad. 
This (2112) pattern starts to 
break up as the energy moves away from the AFM limit;
even for large $S$ ($30$ or $65$), this breakup happens already around
the polyad with $P=10$, so it is much harder than in the FM case to
ascertain the asymptotic pattern of subclustering.
We conjecture that the breakup may happen near the energies 
where,  classically, the stable periodic orbits bifurcate and a 
small bit of phase space goes chaotic.

The barrier for tunneling between the disconnected energy surfaces
has the energy scale of the bare Hamiltonian, which is much larger
(at least, for small $P$) than the scale of effective Hamiltonian $I$ which 
provides the barrier for tunneling among the states in a subcluster.
Hence, the hyperfine splittings are tiny compared to the fine splittings
discussed at the end of Subsection~\ref{subsec:polyads}. 
To analyze numerical results, we replace a degenerate level pair 
by one level and a hyperfine-split pair by the mean level, 
and treat the result as the levels from 
{\it one} of the two disconnected polyad phase spheres,
neglecting tunneling to the other one. 

%%% AFM
Then in the AFM limit, the ``pole'' points in 
Figure~\ref{fig:flows} again 
correspond to spin waves propagating around the ring, while
the stable and unstable points
on the equator are (c) and (b) on Figure~\ref{fig:tracks}.
The three-phase orbit is the {\it highest} frequency orbit in the
AFM limit,~\cite{Houle98c} so again the twofold
and threefold subclusters should occur at the high and low energy ends of
each polyad cluster.
What we observe, however, is that {\it all} the subclusters
are twofold, except the lowest one is often threefold.

% Intriguingly, Figure~\ref{fig:flows} describes the phase sphere of 
% another problem with a cluster of {\it four} 
% spins (or four sublattices of spins) coupled 
% antiferromagnetically (Ref.~\onlinecite{Henley98}. 
% The phase sphere in that case is Fig. 1 of 
% Ref.~\onlinecite{Henley98},
% which looks different until one remembers that in that case, all
% points related by twofold rotations around the $x$, $y$, or $z$
% axes are to be identified. Consequently the sphere of 
% Ref.~\onlinecite{Henley98} has just two distinct octants, corresponding
% to the ``northern'' and ``southern'' hemispheres of 
% Fig.~\ref{fig:flows} in this paper. 
% The special points labeled $T_{\pm}$, $C_{x,y,z}$, and
% $S_{a,b,c}$ in Ref.~\onlinecite{Henley98},
% correspond respectively to the two ``poles'', plus the three stable and
% three unstable points on the ``equator'', in Fig.~\ref{fig:flows}.}

% Note that each main cluster of levels in the present problem 
% corresponds to {\it all} singlet levels for a given spin
% length in the 4-spin problem of Ref. \cite{Henley98}.}

\subsubsection{Isotropic limit}

This regime will includes only $S_{\rm tot} \leq S$
i.e. $E<-1$ -- the same regime in which the flat DOS was observed
(Sec.~\ref{sec-flatspot}). 
Above the critical value $E=-1$, the levels behave as 
in the ``FM limit'' described above. 

At $\sigma = 0$, it is well-known that 
the quantum Hamiltonian reduces to 
${\scriptstyle {1\over 2}}[S_{\rm tot}^2 - 3S(S+1)]$.
Thus each level has degeneracy 
$P \equiv 2S_{\rm tot}+1$. 
(That is the number of ways three spins $S$
may be added to make total spin $S_{\rm tot}$, and each such multiplet
has one state with $S^z_{\rm tot}=0$.)
When $\sigma$ is small, these levels split and will be 
called a polyad.~\cite{FN-isotropicP}

%%% which is a critical value  in
%%% the isotropic limit; 

Classically, at $\sigma=0$
the spins simply precess rigidly 
around the total spin vector. 
%% (The total spin vector lies in the easy plane, since
%% total $S_z$ was constrained to be zero.)
These are harmonic motions of four coordinates;
hence the polyad phase sphere can be constructed by 
(\ref{eq:polyad}). From the threefold symmetry, 
there should again be three orbit types
as represented generically by Figure~\ref{fig:flows}
and Eq.~(\ref{eq:I}). 
For example, an umbrella-like  configuration in which 
the three spin directions are equally tilted out of their plane
corresponds to a three-phase type orbit, with two  cases
depending on the handedness of the arrangement. 
A configuration where
one spin is parallel/antiparallel to the net moment,
and the other two spins offset symmetrically from it),  
follows one of the threefold degenerate orbits. 

Numerically, the level behavior in the near-isotropic limit
is similar to the near-FM limit. 
The fine structure degeneracies are a repeat of the (2112) 
pattern as in the other regimes; the lowest levels of any polyad 
always begin with (1221).
The fraction of threefold subclusters is large here and, 
as expected, grows with $S$, 
(from 0.5 to 0.7 in the case $S=15$). 
However, the energy scales of 
$\alpha P^2$ and $\beta P^3$ behave numerically as $\sigma P^0$ and 
$\sigma P^1$. 
What is different about the isotropic limit is that
the precession frequency -- hence the oscillator frequency $\omega$ --
is not a constant, but is proportional to $S_{\rm tot}$. 
Since perturbation techniques~\cite{FN-I}
give formulas for $I$ with inverse powers of $\omega$, it
is plausible that 
$\alpha$ and $\beta$ in (\ref{eq:I}) include factors of $P^{-2}$ here,
which were absent in the other two regimes.

\section {Conclusion and summary}

To summarize, 
by using detailed knowledge of the classical mechanics of a three
spin cluster \cite{Houle98c},  we have studied the semiclassical
limit of spin in three ways.  First,  using autoregressive
spectral analysis,  we identified the oscillating contributions that
the fundamental orbits of the cluster make to the density of states,
in fact,  we detected the quantum signature of the orbits before
discovering them.  Secondly,  we verified that the quantum DOS
is proportional to the area of the energy surface;
%%  and thus observe
%% the quantum manifestation of topology changes of the classical
%% energy surface as a function of energy.
we also observed kinks in the smoothed quantum DOS,  
which are 
the quantum manifestation of topology changes of the classical
energy surface;
such topology changes, we expect, are more common in spin systems
than particle phase space, since even a single spin has a
nontrivial topology. 
Finally, we have identified three regimes of near-regular
behavior in which the levels are clustered according to a
four-level hierarchy, and we explained many features
qualitatively in terms of a reduced, one degree-of-freedom
system. This system appears promising for two extensions
analgous to Ref.~\onlinecite{Henley98}: tunnel amplitudes
(and their topological phases)
could be computed more explicitly; 
also, 
the low-energy levels from exact diagonalization 
of a finite piece of the anisotropic-exchange
antiferromagnet on the triangular lattice could probably
be mapped to three large spins and analyzed in the fashion
sketched above in Sec.~\ref{sec-clustering}. 

\acknowledgments

This work was funded by NSF Grant
DMR-9612304, using computer facilities of the Cornell Center
for Materials Research supported by NSF grant DMR-9632275.  We 
thank Masa Tsuchiya,  Greg Ezra,  Dimitri Garanin,  Klaus Richter and
Martin Sieber for useful discussions.

%% \bibliographystyle{prsty}
%% \bibliography{journals,misc,classical,semiclassics,magnetism,math,anons,informatics}

\begin{thebibliography}{10}
\bibitem[\dag]{clhnote} 
Author to whom correspondence should be addressed.
\bibitem{FN-maslov}
$A_p(E)$ contains a phase factor,
$e^{i\mu/4}$,  where $\mu$ is the Maslov index
and depends on the topology of the linearized
dynamics near the orbit;  see 
J.~M. Robbins,
Nonlinearity {\bf 4}, 343 (1991).
As we do not consider amplitude or phase,  the
Maslov index is irrelevant for this paper.
\bibitem{FN-FTnote}
Eq. (\ref{eq:ftospec}) is for illustration.  The square window
aggravates artifacts of the Fourier transform which could be reduced by
using a different window function (See Ref.~\onlinecite{Marple87}).
\bibitem{FN-Paxes}
Observe that the ``north pole'' symmetry axis in Figure~\ref{fig:flows}
is the $z$ axis. On the other hand, the axis of the polar coordinates
$(P_x,\Psi_x)$ in Eq.~(\ref{eq:polyad}) is the $x$ axis and passes 
through two stationary points of types (b) and (c) in the figure. 
The equator in Figure~\ref{fig:flows} includes all motions 
with $\Psi_x=0$ or $\pi$, i.e. the oscillations are in phase.
\bibitem{FN-isotropicP}
In terms of the two oscillators, $P=2S+1$ implies that
polyads are allowed only with a net even number of quanta;
we do not yet understand this constraint.
\end{thebibliography}

%%% Local Variables: 
%%% mode: latex
%%% TeX-master: "document"
%%% End: 

\end{document}